# Laser Structured Optical Interposer for Ultra-dense Vertical Coupling of Multi-core Fibers to Silicon Photonic Chip

Gligor Djogo, Amir Rahimnouri, and Peter R. Herman

*Abstract*—Femtosecond laser writing in glass was harnessed to fabricate photonic circuits comprised of waveguides, micro-mirrors, and fiber sockets, in an optical interposer chip. The compact design enabled ultra-dense routing of 40 optical channels from six multi-core fibers, in a two-layer array, vertically coupling onto a 2D grid of silicon photonic grating couplers. Various photonic circuit designs were explored and tested, achieving an average single-pass insertion loss of -5.0 dB for the fully packaged system, with a minimum loss of -3.2 dB. The reasonably low loss and low-profile, compact package make such interposers an attractive option for densifying interconnects to address bottlenecks in optical networks, datacenters, and optical computing applications.

## I. INTRODUCTION

SILICON photonics (SiP) has developed into a powerful integrated optics platform for telecommunications, datacenters, quantum optics, and biomedical applications [1,2,3 Ghiasi, Yu, Siew]. In comparison to the miniaturized SiP chips, the advanced optical networks used to interconnect between chip devices are based on optical fiber, harboring waveguides with dimensions about 30-fold larger. Bridging between these dissimilar platforms has been a major challenge in next generation optical interconnects that are aiming for higher channel-densities and scaling up of channel counts, all the while maintaining low-loss designs [4 Marchetti].

Edge coupling of single-mode fibers directly on the chip facet is a well-established technique, where engineered tapers and lensed fibers have overcome large mode-mismatch loss between the fiber and the chip waveguides [4,5 Marchetti, Nauriyal]. Limits in the linear packaging density of fiber arrays can be overcome with 3D waveguide fanouts formed in glass interposers [6,7,8 DjogoIJEM, Zhao, Doany], facilitating high packing densities of up to 50 channels/mm in the SiP edge when routing from similarly dense multi-core fibers (MCF). These channel spacings are at the threshold of crosstalk in silica-based waveguides, thus limiting the 1D scalability for edge coupling.

To better harness the high areal density of 448 cores/mm$^2$ available today in MCF [9 VIKopp], one must consider vertical coupling to the surface of the SiP chip [4 Marchetti]. Vertical coupling may be driven either through waveguide-to-waveguide adiabatic coupling [10,11 Brusberg, Poulopoulos] or grating coupling [12,13 CKopp, Mirshafiei]. The latter is preferred due to more relaxed alignment and packaging tolerances, as well as higher coupling efficiencies by mode-matching of the silicon-grating to fiber waveguide modes. These advantages tend to outweigh the restrictions on the bandwidth (i.e., < 100 nm) and the selective polarization response of the gratings [12,4 CKopp, Marchetti]. Simple packaging of fiber orthogonally onto a 2D grid of grating couplers [9,14 VIKopp, Ding] has been demonstrated up to 37 fiber channels, while a more robust packaging solution used an interposer with a waveguide fanout impinging vertically on SiP grating couplers arrayed at 87 channels/mm$^2$ density [7 Zhao]. However, the optical fibers remained vertically mounted above the SiP chip owing to the limited bend radius available.

Facilitating dense, 3D routing of channels in vertical coupling requires new means for creating steep waveguide bends within the interposer, to enable a flattening of these intermediate coupling layers into compact packages. Alternatively, embedded mirrors within the interposer can offer low-profile packaging onto the surface of a SiP chip that meet robustness and compactness requirements for industrial settings. In one approach, mirrored surfaces, often based on total internal reflection (TIR), have been proposed in a variety of geometric configurations [13,15,16,17,18,19 Mirshafiei, Wang, Cheng, Huang, Boucaud, Chou], where machining, etching, or polishing away of substrate material enables tight redirection of light from inside fiber or within optical circuits. In glass, hollowed-out trenches have shown to be strong TIR reflectors [15,16 Wang, Cheng], including fiber-to-waveguide

Manuscript received XX Month 2024; revised XX Month 2024; accepted XX Month 2024. This work was supported by the Natural Sciences and Research Council of Canada (NSERC) under grant XXX. *(Corresponding author: Gligor Djogo).*

The authors are with the Edward S. Rogers Sr. Department of Electrical and Computer Engineering, University of Toronto, Toronto, ON M5S 1A4, Canada (e-mail: g.djogo@mail.utoronto.ca).

Color versions of one or more of the figures in this article are available online at http://ieeexplore.ieee.org

Digital Object Identifier XXX.

coupling [17 Huang], as long as the surface smoothness is minimized. However, such macro-scale mirror-waveguide circuits have failed to meet the higher channel densities suitable for optical interconnects in SiP chips.

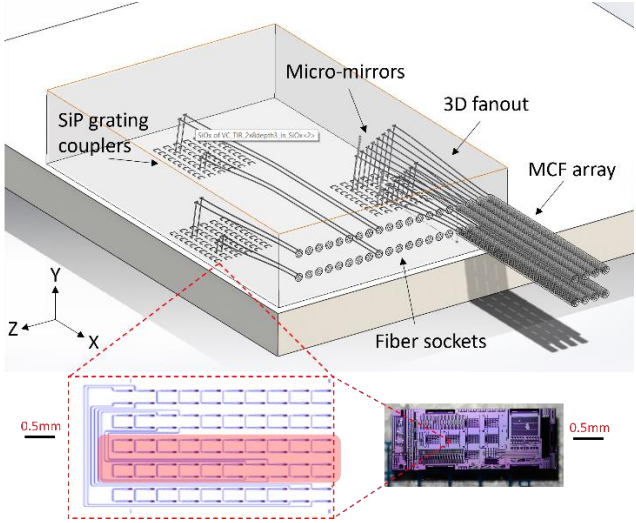

**Fig. 1.** Vision for a high-density vertical coupling platform using micro-mirror and 3D waveguide circuits to interconnect MCFs to SiP gratings. Insets show the grating layout on a real SiP test chip, with the highlighted loops being targeted in the paper.

Microfabrication techniques in polymer interposers have permitted waveguides from a single horizontal plane to terminate in TIR wedges, and couple vertically into SiP [18,19 Boucaud, Chou]. Extension to 3D fanouts would be necessary to fully exploit the high areal density available from grating couplers. Glass interposers would offer higher robustness and thermally stability over polymer materials. To this end, femtosecond laser irradiation with chemical etching (FLICE) [20,21 Hnatovsky, Marcinkevicius] is a promising direction for opening micro-disks at arbitrary 3D positions in a glass substrate and serve as TIR mirrors. FLICE surfaces with smoothness down to 10 nm rms have been demonstrated [22 Ho]. Such optical-grade disks have been applied as micro-optical resonators within a 3D waveguide circuit [23 Haque] and as micro-mirror reflectors for high angled turns of buried waveguides [24 Amir?]. The facility of 3D writing to precisely register multi-level TIR optical components with 3D waveguides may thus be extended over a large glass volume, as waveguides can be routed around, above, or below mirrors to generate compact interposer designs for vertical coupling to SiP chips, without trading off system performance.

In this paper, a fully packaged glass interposer is presented between a 2D array of MCFs and grating couplers on a SiP chip, demonstrating low-loss vertical chip coupling through 40 waveguide channels. Femtosecond laser micro-structuring was harnessed to create high-density waveguide circuits and etched TIR micro-mirrors on multiple layers inside the interposer, as shown schematically in Fig. 1. The design also incorporated fiber sockets [6,25 DjogoIJEM, DjogoJLT] that facilitated self-alignment of MCFs into the interposer, while 3D laser writing permitted precise registration of the waveguide circuits to each optical element with minimal added loss. The optical coupling through the interposer was limited by the bandwidth and polarization sensitivity of the SiP gratings. Thus, the fully packaged design demonstrates a step towards scaling up robust, low-profile, high channel-density coupling systems, reaching towards the large-scale coupling vision laid out in Fig. 1.

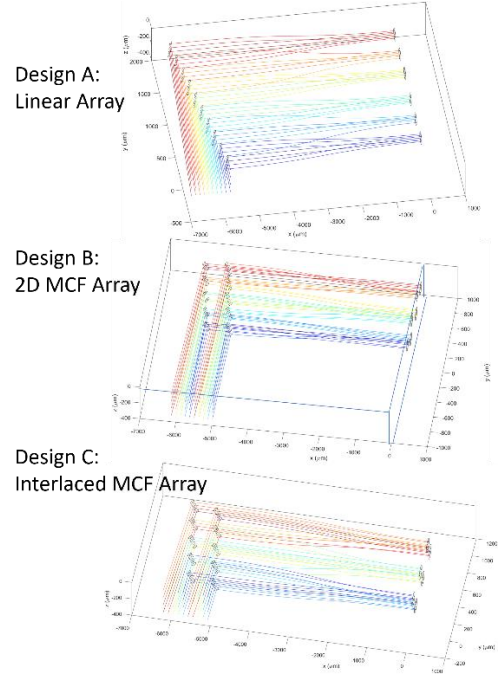

**Fig. 2.** Waveguide paths and micro-mirror arrangements traced out for the three circuit designs. (a) Design A spreads out ten mirror groups in a linear array, (b) Design B compresses the mirror arrangement by introducing additional S-bends, and (c) Design C achieves a compressed arrangement with waveguide intersects instead.

## II. INTERPOSER DESIGN

The interposer design targeted a three-dimensional matching of waveguide links between on-chip grating couplers (see [7 Zhao]) laid out at a density of 10 grating couplers per row on 127 μm by 90 μm pitch (Fig. 1 inset) and multicore fibers. The present MCF (Chiral Photonics, MCF-007_2) provided six waveguide cores arranged in a hexagonal pattern around a seventh central core, with 37 μm core-to-core separations. In prior work, this MCF had been packaged with a 3D interposer at a conservative linear density of 250 μm fiber-to-fiber spacing, facilitated by a novel approach using self-aligning sockets [6 DjogoIJEM]. For the present purpose, such a single layer MCF array (28 core/mm) fell about 3-fold short of matching the grating-coupler density (i.e., 111 couplers/mm) on the SiP chip. Two versions of denser fiber socket coupling were thus targeted, packaged in one row of six stacked MCFs (Design A) or two rows of three MCFs (Design B and C), as shown in Fig. 2.

In total, the interconnection of 40 input/output channels over four rows of grating couplers (x = 1.143 mm by y =

0.27 mm) were tested in three versions of micro-mirror layouts (Fig. 2). The 4×10 section of grating couplers had SiP loops between nearest neighbors, that yielded a -4.9 dB insertion loss for TE polarized light near the 1550 nm telecommunication band when probed by SMFs.

The three designs of micro-mirror arrangements showed a progression of mirror-waveguide circuitry design from least to most compact. Design A was based on the simplest waveguide routing configuration, leading to TIR mirrors organized into ten groups of four, and spread over an interposer height of 2 mm. Designs B and C compressed the mirror arrangement 2-fold, to 1 mm, by reconfiguring five groups of mirrors to align at similar planes as the lowest 5 mirror groups in Design A. This arrangement pushed pairs of mirrors onto overlapping waveguide tracks for coupling with the MCF, requiring a re-routing of waveguides and/or offsetting of mirror positions to enable waveguides to bypass the first set of mirrors. In Design B, the back group of mirrors were offset laterally by 30 μm in ±Z (Fig. 2(b)) while in Design C the back group of mirrors were sifted in 50 μm in +Y (Fig. 2(c)).

In Designs A and C, an array of 10 straight waveguides were formed at 10° angle off vertical and arranged in 4 vertical planes to match with the 40 grating coupler positions and coupling angle. In Design B, S-bends were required to match grating positions with the 30 μm offset in mirror positions (±Y), adding 1 mm to the interposer height, and thus matching the 2 mm thickness in Design A. In order to maintain the 1 mm interposer thickness in Design C, many of the vertical and horizontal waveguide paths were required to intersect at a large 80° angle. This angle was found to pose a minimal factor in insertion loss and crosstalk in the interposer.

From the various micro-mirror arrangements, 3D waveguide fanouts were laid out with MCF positions that minimized net bending loss. The minimum bend radius was restricted to 50 mm [6 DjogoIJEM] and vertical and horizontal bends were combined into a single S-bend along a tilted plane, further shortening the fanout length. Overall losses were also minimized by varying bending radii on individual channels to maintain a minimum waveguide-to-waveguide distance of 28 μm in all design variations of the fanout section, limiting the potential for channel crosstalk. The length of fanout on all three designs in Fig. 2 ranged from 4 to 6 mm (in X).

Waveguide fanout losses were simulated for various combinations of mirror-to-fiber core mappings, where 40 mirror positions in Designs A, B, and C were directed at 6 MCF positions. Vertical and horizontal components of bending loss were assessed from prior work [6,25 DjogoIJEM, DjogoJLT]. The MCF positions were iteratively tuned to minimum loss designs that were presented in Fig. 2.

## III. Methods

To fabricate the interposer, a laser microfabrication system was used to irradiate a silica glass wafer (PG&O, Corning 7980 polished fused silica?) of 1 mm thickness and inscribe the waveguide circuitry and various components. A fiber laser (Amplitude, Satsuma HP[2]) was frequency doubled, producing about 250 fs duration pulses at a 500 kHz repetition rate and electronically controlled pulse energy. The laser beam was directed into the wafer and focused to a $\omega_o = 0.8$ μm radius ($1/e^2$ intensity) with a 40X aspheric air lens (Newport, 5722-A-H) to induce modifications in the glass material. The sample was mounted on computer-controlled precision motion stages (Aerotech, PlanarDL-200XY and ANT130-110-L-Z), integrated with the laser system, giving submicron precision and 200 nm repeatability when positioning the beam focus.

TABLE I
SUMMARY OF LASER STRUCTURING PARAMETERS FOR MIRROR-WAVEGUIDE CIRCUITS

| Parameter | Loss Estimate |
|---|---|
| Waveguide propagation loss | |
| 50 μm – 168 nJ | 0.20 dB/cm |
| 140 μm – 148 nJ | 0.56 dB/cm |
| 230 μm – 148 nJ | 0.65 dB/cm |
| 320 μm – 128 nJ | 0.71 dB/cm |
| Waveguide coupling loss | 0.3 dB/end |
| S-bend loss (R=50mm) | |
| Horizontal bend average | 0.5 dB for 100 μm offset |
| Vertical bend average | 0.4 dB for 50 μm offset |
| Additional intersect loss (100°) | < 0.01 dB/intersect |
| Additional socket coupling loss | < 0.2 dB |
| Additional MCF core misalignment loss | < 0.2 dB |
| TIR mirror reflection loss | 0.4 to 0.8 dB |

Waveguide paths were formed of a single continuous laser scan at 0.1 mm/s, with the laser polarization parallel to the scanning direction. The interposer design required 3D paths extending over a depth range from 50 μm to 320 μm (Y axis in Fig. 1). When focusing deeper into the wafer, a moderate influence of increasing aberration from the air-glass surface weakened the waveguide, requiring a means of depth compensation. Waveguide loss optimizations for straight and S-bend paths was carried out using standard techniques of fiber probing as previously reported [6 DjogoIJEM].

Amongst a variety of laser exposure conditions, select parameters were identified for low-loss waveguide writing across depths at 1550 nm wavelength, as summarized in Table I. The propagation losses were calculated from measured insertion losses on 0.5-inch-long samples, and reduced by coupling losses of 0.3 dB as inferred from calculation of the modal mismatch between the waveguide and probing SMF. The reported losses concur with values in previous work [6,25 DjogoIJEM, DjogoJLT].

The optimal exposure parameters for writing straight waveguides were further extended to horizontal and vertical S-bends with decreasing radii. The additional loss due to the S-bends spanning laterally by 100 μm for horizontal and spanning 50 μm for vertical bends were assessed around the four main writing depths (50, 140, 230, and 320 µm). As in past work, a minimum bend radius of 50 mm was selected for both horizontal and vertical bends, yielding average bend losses of 0.5 dB and 0.4 dB respectively when averaged across depths, as given in Table I.

Losses and channel crosstalk arising from the intersections of straight waveguides were further evaluated for crossing angle of 80°. No significant additional loss was incurred with up to 16 waveguide crosses, yielding at most 0.01 dB per intersect as noted Table I. The crosstalk was evaluated by connecting parallel straight waveguides with one or multiple intersecting waveguides and resulted in no measurable crosstalk above the 40 dB noise floor (Table I).

The outer structures for TIR micro-mirrors and fiber mounting sockets were traced with overlapping laser scans, with laser polarization aligned to assemble the volume nanogratings with their surfaces parallel to the optically critical surfaces of the reflective micro-disk and the back face of the socket. Thus, near optically smooth walls (i.e., $\sigma_{rms}$ = 10 nm [22 Ho]) were generated at these locations with laser energy of 40 nJ/pulse and 0.4 mm/s scanning speed, leading to low mirror insertion losses of at most 0.8 dB for waveguide-to-waveguide coupling [24 Amir?] and less than 0.2 dB added loss when fiber coupling to waveguides through the socket [25 DjogoJLT], as summarized in Table I. The laser tracks for the mirror access ports were optimized for rapid acid etching by using higher exposures of around 200 nJ/pulse and 0.4 mm/s scanning speed.

The micro-mirrors were tilted at 50° with respect to the waveguides to induce TIR. The elliptical disks provided a reflecting surface of 30 µm tall and ~54 µm wide to encompass the diffracted waveguide mode size and provide a buffer to surface imperfections on the disk perimeter. Following prior work [24 Amir?], the lengths of access ports were tailored to equalize etching times across the four principal mirror depths in the glass and thus ensure a minimal surface roughness across all 40 mirrors.

After laser inscription, the wafer was immersed in 5% hydrofluoric acid to complete the FLICE structuring process of the glass following prior procedures [6,22,23 DjogoIJEM, Ho, Haque]. The etching bath was stirred at a moderate 60 rpm rate, and etching time was tuned in 5-minute intervals following an initial 3-hour bath. Fig. 3 shows backlit microscope images of fully etched interposer circuits for designs A, B, and C, but without fiber alignment sockets. Full etching was noted when the mirror disks turned dark, owing to TIR from the near planar surfaces, with uniformly shaped structures over all depth positions.

The same FLICE procedure was also applied to open self-alignment sockets in the interposer facet, improving over our prior work in [6,25 DjogoIJEM, DjogoJLT]. The socket length was increased from 50 µm to 250 µm to improve the ruggedness of the fiber bonding, while also matching the deep mirror etching time of about 3 hours. Socket positions were aligned in 1 by 6 or 2 by 3 configurations according to the designs in Fig. 2. The two-layer sockets facilitated a denser 2D packing of the MCF array on center-to-center socket spacings of less than 300 µm, yielding a high areal packaging density of 133 waveguide channels per square millimeter. Socket parameters were tuned independently at the two depths, to provide well-sized sockets enabling micron-level alignment of MCF cores to interposer waveguides. Different laser pulse energies were used to write the socket cylinder, 80 nJ deep compared to 100 nJ when shallow, the deeper socket diameter was reduced by 2 µm to match MCF once etched, and the socket body was shifted by 1 µm (+Z) for shallow sockets and 3 µm (+Z) for deep sockets, to center the MCF on the waveguides. A larger gap of 11 µm was also needed to prevent etching from the backplane into fanout waveguides.

(a) Design A: linear array

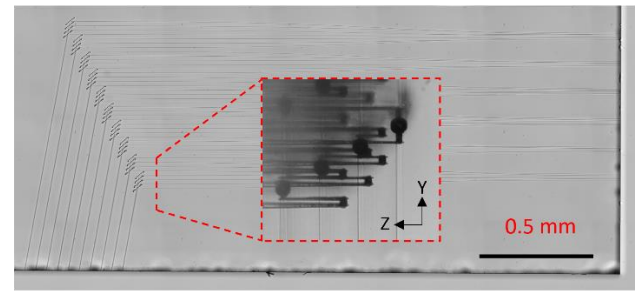

(b) Design B: 2D MCF array

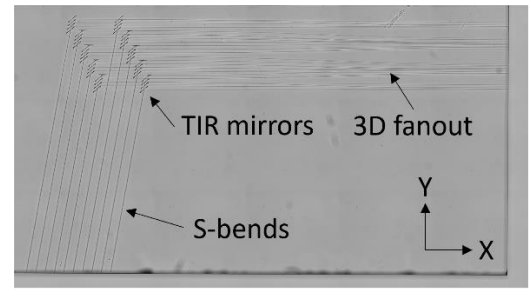

(c) Design C: interlaced 2D MCF

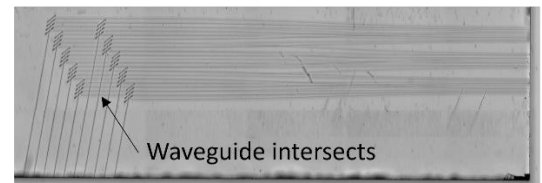

**Fig. 3.** Composite images of fully etched micro-mirror and waveguide circuits for (a) Design A, (b) Design B, and (c) Design C. The inset shows an end view of the four micro-mirrors forming a group, positioned at 50, 140, 230, and 320 µm below the glass surface (left edge of image).

First-level analysis of the interposer circuits was accomplished by direct SMF fiber probing of waveguide channels through butt-coupling with index matched oil (Cargille, Refractive Index Liquid). The interposer samples and fibers were held in place and positioned on multi-axis alignment stages (Luminos, Cor/Align P4, P5, and P6), tilting the SMF probe to 10° on the grating coupler side. Broadband light from a multi-diode source (Agilent, 83437A) and optical spectrum analyzer (Ando, AQ6317B) was used to record insertion loss spectra across the telecomm bands. Once MCF sockets were added to the interposer, a commercial MCF fanout (Chiral Photonics, MCFFO-P-07/37-1550-SM01-FC/APC-60) was applied to probe the 7-channel groups of waveguides in parallel, following similar butt-coupling and index matching techniques but keeping the socket side free of oil. Measured interposer losses were adjusted to account for the MCF fanout device average insertion loss of 1.0 dB (standard deviation $\sigma_{rms}$ = 0.1 dB). This MCF device was also used in the interposer packaging process, as summarized previously [6,25 DjogoIJEM, DjogoJLT]. Epoxy optical adhesive (Norland, NOA81) was dispensed onto the distal fiber end and a simple plug-and-twist alignment of the MCF into the socket was followed with UV lamp curing (EXFO, Novacure 2100). The long sockets (250 μm) offered strong fiber grip without the need for excess adhesive application to ensure long-term bonding stability.

The interposer-to-chip packaging followed similar protocols of active alignment, finishing with UV curing of a thin index-mated epoxy layer, as summarized in past edge coupling work [6 DjogoIJEM]. Interposer channel losses were assessed before and after the two chips were permanently epoxy bonded. Fig. 4 shows an image of the fully packaged interposer, epoxy bonded to the SiP chip, and fiber packaged with the assistance of the self-aligning sockets (inset image). The single-pass loss was calculated by measuring the total loss through a pair of waveguide channels, connected by a SiP grating loop, and dividing by two after removing losses from the grating loop and MCF fanout probe (repeated later??). Due to the SiP grating layout (Fig. 1) only 18 short loops were addressed in the final packaged chip-to-chip alignment. However, all channel pairs were tested on short loops before final packaging, by repositioning the interposer. In this way, two long loops were probed spanning approximately 575 μm diagonally across the SiP chip.

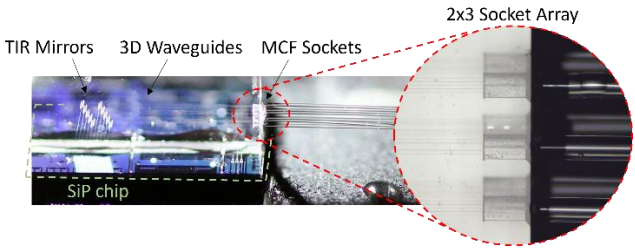

**Fig. 4.** Images of the fully packaged optical interposer for vertical coupling of MCF (right) to SiP chip (bottom). The inset shows the 2D array of self-aligning fiber sockets used to facilitate high-density packaging.

## IV. RESULTS AND DISCUSSION

For each of the three interposer designs presented in Section 2, three versions of waveguide mapping were assessed for propagation losses between the micro-mirrors and MCF sockets by applying the loss parameters in Table I. Results for average fanout losses as estimated over all 40 channels are reported in Table II (column 3). Higher loss was found to correlate with larger lateral offset of the waveguides (column 2). For example, in Design B, smallest waveguide offsets averaging in a range of 65 to 69 μm (with root mean square standard deviation $\sigma_{rms}$ = 17 to 36 μm) yielded a minimum overall fanout loss varying from -1.12 to -1.26 dB (Table II, column 3). In contrast, the doubling of the waveguide offsets (103 to 153 μm) in Design A increased losses by ~ 0.4 dB. The micro-mirror reflection and angled vertical waveguide sections added a further loss of ~ 0.9 dB, yielding the full chip losses varying from -2.03 to -2.45 dB (Table II, column 4) over the total of nine circuit designs. Design C provided the most compact interposer profile, yielding similar low overall losses to Design B. Only minor variances in losses were noted in the different waveguide mappings, offering design flexibility in the channel-to-channel routing. Designs A-3, B-2, and C-1 (Fig. 3) were selected for laser fabrication.

TABLE II
CHANNEL LOSS CALCULATED FOR DESIGN A, B, AND C WITH VARIOUS MCF-TO-MIRROR MAPPING OF WAVEGUIDES

| Design – Mapping | Fanout offset Avg. / St.Dev. | Loss Estimate Fanout Avg. | Loss Estimate Full Circuit Avg. |
|---|---|---|---|
| A – 1 | 153 ± 69 μm | -1.65 dB | -2.45 dB |
| A – 2 | 142 ± 68 μm | -1.65 dB | -2.45 dB |
| A – 3 | 103 ± 33 μm | -1.54 dB | -2.34 dB |
| B – 1 | 67 ± 17 μm | -1.18 dB | -2.09 dB |
| B – 2 | 65 ± 22 μm | -1.12 dB | -2.03 dB |
| B – 3 | 69 ± 36 μm | -1.26 dB | -2.18 dB |
| C – 1 | 71 ± 27 μm | -1.20 dB | -2.06 dB |
| C – 2 | 71 ± 24 μm | -1.20 dB | -2.06 dB |
| C – 3 | 76 ± 30 μm | -1.33 dB | -2.19 dB |

Fig. 5 summarizes the results of insertion loss measured at 1550 nm wavelength across all 40 channels in each of the interposer designs, comparing the fanout section (blue) with the full chip losses (black). The full chip combines the fanout with mirror and angled waveguides but excludes the fiber self-aligning sockets. A cyclical pattern of loss in the data follows a near-periodic variance in guiding layer depths (4 layers) and S-bend offsets, with increasing channel number. In Design A-3 (Fig. 5(a)), a repeating progression from shallow to deeper waveguides on 4 channel cycles manifested in losses cycling between -1.4 to -1.9 dB, with an average of -1.62 dB and standard deviation of 0.14 dB for the fanout-only (blue) interposer. The 0.5 dB higher loss for the deeper waveguides arises from surface aberration effects on deep and large NA laser focusing as widely reported in the literature [26 Ehsan, others]. The addition of TIR mirrors and angled vertical waveguides can be seen to disturb the cyclical pattern and add a further ~ 1 dB increase to the insertion losses to yield an average -2.63 dB loss ($\sigma$ = 0.24 dB). The higher number channels with higher reported losses (~ 2.9 dB) correlated with the longest waveguides in the vertical coupling section of the circuits as seen in Fig. 3(a).

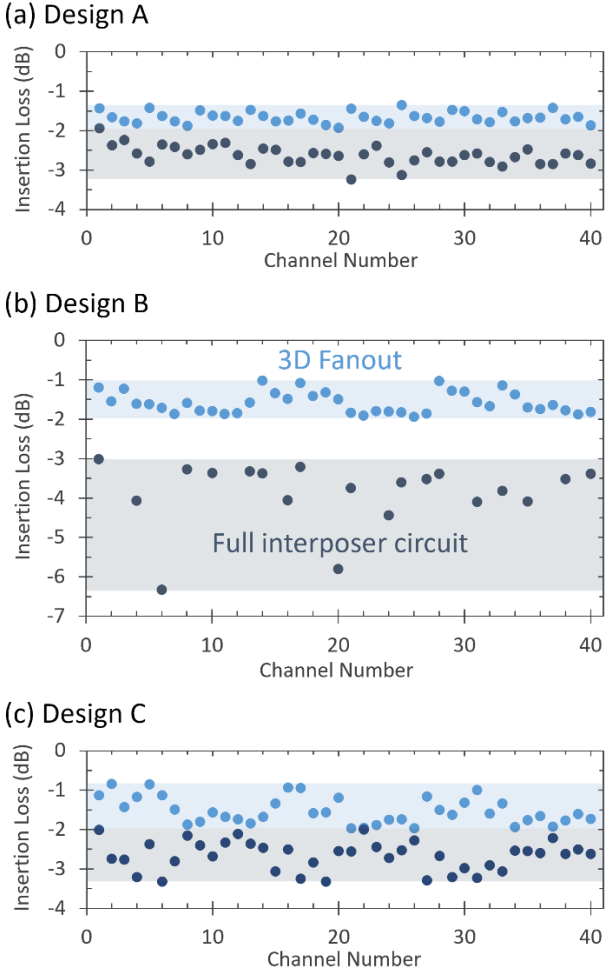

**Fig. 5.** Insertion losses of fanout section (blue) and full interposer circuit (black) as measured by SMF probing across 40 interposer channels for the three micro-mirror arrangements of (a) Design A-3, (b) Design B-2, and (c) Design C-1. Fiber alignment sockets were excluded in this prototype.

The more varied MCF-to-mirror mappings in Design B and Design C (Fig. 5(b) and 5(c)) were less regular in patterning, correlating with both the 4-cycle depth-dependent loss and the S-bend offset loss. In Design B and C, fanout losses followed the MCF groupings, with the three shallower MCFs in the 2D array yielding generally lower losses due to lower surface aberration. Design B provided an average insertion loss of -1.58 dB ($\sigma$ = 0.27 dB) in the fanout, which increased to -3.87 dB ($\sigma$ = 0.82 dB) in the full-circuit interposer. In comparison, Design C provided an average insertion loss of -1.52 dB ($\sigma$ = 0.33 dB) and -2.67 dB ($\sigma$ = 0.53 dB) for the fanout and full-circuit interposers, respectively.

Overall, the experimental loss data could be well represented by the simulated results in Table II, particularly for Design A-3 where measured fanout and full interposer losses were 0.1 dB higher and 0.3 dB higher, respectively, than the estimates. Larger discrepancies were noted in Designs B and C, yielding respective losses that were 0.5 dB and 0.3 dB higher in the fanout, and 1.8 dB and 0.6 dB higher in the full interposer. The increased losses in these designs may arise from effects such as a directional writing dependence of the S-bends in the vertical coupling waveguides (Design B), a detuning of optimal mirror depths by ± 30 μm (Design B), coupling losses due to proximity of waveguides to the mirror etching ports (Designs B and C), and intersections of waveguides (Design C). Nevertheless, the overall good correspondence of the measured loss data demonstrated the reproducibility and routing flexibility in scaling up laser writing of optical circuits over 3D volumes.

The analysis of the three prototype interposers favored the selection of Design C-1 for chip-to-chip packaging on the merits of interposer compactness and low insertion loss. Fiber sockets were included, with modifications described above in the methods section. An additional tuning of laser-written waveguide depths was necessary to improve alignment of the guided mode positions with the SiP grating array positions and the MCF core positions. Vertical writing depths were shifted by up to 10 μm (+Z) to compensate for the underlying surface aberration and nonlinear propagation effects. Laser track depths of 50, 136, 223, and 310 μm minimized the coupling loss simultaneously between all four grating layers within a processing reproducibility of ± 1 μm on depth. The coupling to the MCF cores was also found to improve by increasing the hexagonal separation of laser written waveguides to 38.5 μm (from the 37 μm design).

The channel-by-channel insertion loss of the interposer is shown in Fig. 6(a) to offer a low average loss of -2.75 dB with a small spread of $\sigma$ = 0.96 dB (standard deviation). Fig. 6(b) shows the interposer channel response is relatively flat over a wide spectrum, for example, ranging between -2.9 to -4.0 dB across a source bandwidth of 420 nm for channel 8. Fig. 6(b) also presents the full loop loss spectra for the fully packaged interposer (Fig. 4), showing a similar loss for a short loop channel (loop 11) versus a long loop channel (long loop 1). The full loop loss has climbed sharply to over -16 dB, while also narrowing significantly to 48 nm bandwidth (3 dB) with a peak wavelength near 1504 nm. The major contribution to the spectral narrowing and increased loss was attributed to the grating coupler, which yielded a minimum 4.9 dB loss (double pass) on testing with direct SMF probing. The remaining contributions to insertion loss at the 1504 nm peak were ~3 dB polarization loss at the SiP chip, two-pass interposer losses of 3.0 dB (channel 8) and 3.2 dB (channel 10), and a double pass MCF probe loss of ~2 dB.

A full channel assessment of the interposer-SiP package (Fig. 4) is presented in Fig. 6(c), comparing the single-pass loop losses prior to epoxy bonding (pre-curing) with immediately after (packaged). Full loop losses attributed to the gratings (4.9 dB), the polarization (3 dB), and the MCF probes (2 dB) were subtracted, and remaining loss was divided by two. The average single-pass loss for short loops (blue circle) increased marginally from -4.65 dB ($\sigma$ = 0.64 dB) under 'pre-curing' alignment of components to -4.98 dB ($\sigma$ = 0.89 dB) after bonding. The average long loop losses (orange squares) were ~0.4 dB higher, possibly arising from waveguide-to-grating alignment errors compounded over the four grating layers in the SiP chip. The higher 0.3 dB loss during the epoxy bonding step arose in part (0.2 dB) from a random misalignment of ± 1 µm in the waveguide core position along the MCF length, as previously reported [25 DjogoJLT]. Epoxy curing between the interposer and SiP chips further disturbed the waveguide array positioning, leading to either an increase or decrease in losses randomly over the channels. A reassessment of the insertion losses two weeks after packaging yielded an insignificant degradation of less than 0.1 dB, pointing to the long-term stability of the vertical coupling and MCF packaging in the present interposer-to-SiP chip package.

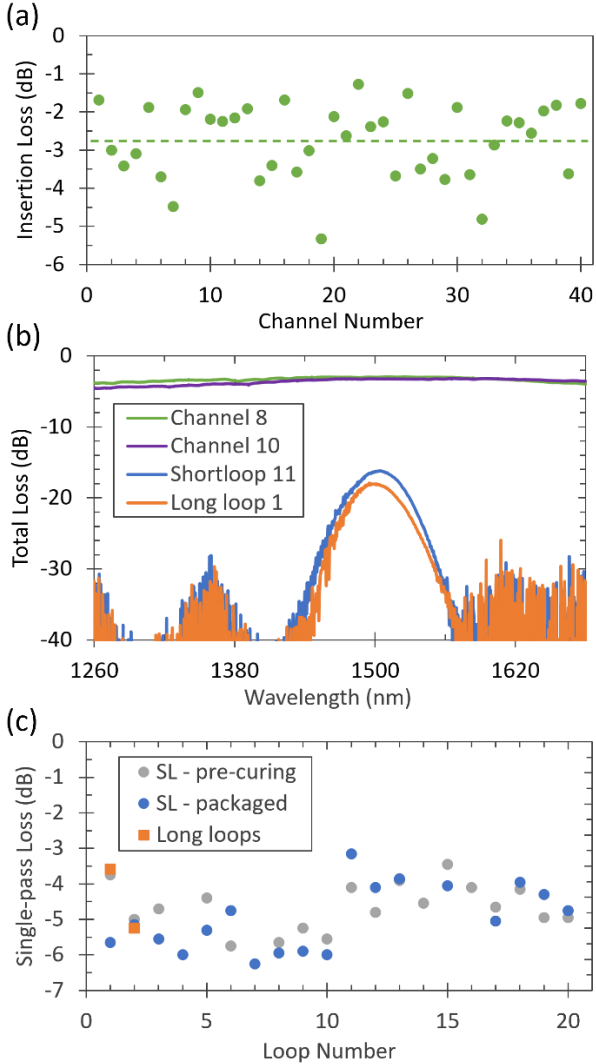

**Fig. 6.** Losses before and after the final photonic packaging of Design C-1 vertical coupling interposer. (a) Insertion losses of the full mirror-waveguide circuit with fiber sockets, probed at 1550 nm guiding. (b) Select insertion loss spectra for two interposer channels and round-trip loss spectra for one short loop and one long loop. (c) Summary of single-pass loop losses during the packaging to SiP chip.

The average single-pass losses of -5.0 dB (Fig. 6(c)) in the vertical coupling interposer compares favorably with the -6.5 dB loss reported for the edge coupling interposer in our previous work [6 DjogoIJEM]. The TIR mirrors in the present interposer facilitated shorter waveguide routing from the MCF to the SiP chip gratings, and the potential for much higher areal channel density (150 channels/mm$^2$) owing to the 2D grating array that was otherwise restricted to the edge coupling case to about 33 channels/mm. Hence, it also achieved the minimum single pass loss of -3.2 dB in loop 11, for both vertical and edge couplers.

In the push to drive up the density for fiber-to-chip interconnection Kopp et al. [9 VIKopp] harnessed a finely pitched tapered fiber array for ultra dense coupling at 1250 channels/mm$^2$ and 3 dB average loss to SiP grating arrays. However, this direct fiber-to-grating coupling presents an intrusive geometry that precludes the benefits of 3D multi-layered packaging as facilitated by the present interposer design (Fig. 4) via a thin (2 mm) silica optical circuit. Dense 2D arrays of MCF (6 MCF at 150 channels/mm$^2$) on the edge of the interposer were interconnected with the dense 2D array of vertical grating couplers (87 channels/mm$^2$) on the SiP chip. The interposer average single-pass loss of -5.0 dB at 1505 nm wavelength (Fig. 6(c)) which improves to approximately -4.0 dB if the losses in both of the waveguide and grating designs were minimized to the same 1550 nm wavelength (Fig. 6(b)).

An interposer design by Zhao et al. [7 Zhao] provided an improved single-pass loss of -2.4 dB at similar high-density design of 87 channels/mm$^2$ to the SiP surface, but without the TIR mirrors. With a comparable footprint of 6 mm by 2 mm, the fibers projected vertically from the interposer standing 19 mm tall above the SiP surface, thus eliminating the potential for multi-level stacking of board level integrated electronics and photonics. Other interposer approaches for low-profile vertical coupling have relied on TIR as induced by angle polished wedges intercepting waveguides. With such wedges, laser written waveguides [13 Mirshafiei] offered -1.7 dB insertion loss, polymer waveguides offered -5.5 dB loss [18 Boucaud], and ion-exchange waveguides offered -1.6 dB average loss [27 Schroder]. In these cases, various limitations were imposed by the large wedge structure and planar waveguide circuitry, precluding a scaling up to the higher channel numbers (40) or channel densities (87 ch/mm$^2$) in the present interposer (Fig. 2), without the need for complex stacking and packaging of the multi-layered wedge devices. For example, Schroder et al. presented a 9-channel interposer with low density of 4 channel/mm$^2$.

Direct writing of polymer waveguides in 3D, known as photonic wire bonding, has coupled MCF and waveguides with low loss and is a rapidly developing technology for high-density interconnects [28 Lindenmann]. However, polymer waveguide materials have not offered the robustness and scalability provided by dielectric waveguide circuits. The selection of the fused silica substrate in this work was driven by the FLICE processing, enabling low-loss TIR micro-mirrors (< -0.8 dB) to be embedded and interfaced with 3D waveguide circuits. The mirror-to-mirror spacing, as small as 50 µm in the present interposer, defines a new approach for high-density 3D waveguide circuitry at a rate of 444 mirrors/mm$^3$ that has not previously been applied in fiber to SiP chip packaging.

With the laser propagation on the z-axis (Fig. 1), the present micro-optic circuit fabrication may be freely extended in the X and Y directions to scale up the I/O channel count with more rows of MCF and matching optical circuit layers. This footprint offers 40 new channels for every millimeter added in interposer height. Extension of the interposer in the Z direction would require laying out a more complex network of TIR etching ports to breach the surface, as well as beam shaping approaches to compensate the deep-writing effects of surface aberration [26 Ehsan].

A large component of the -2.75 dB channel loss in the present interposer (Fig. 6(a)) is due to waveguide propagation loss (value) which may be improved by further tuning of the laser beam delivery, for example, by multi-pass writing [29 AjoyKar] Other compositions of glass, notably Eagle glass [30 Withford], have demonstrated significantly lower waveguide propagation loss, but do not offer the 3D FLICE processing capability for TIR mirror formation. Mirror losses of -0.8 dB may be addressed by development of a thermal annealing process [16 Cheng] or alternate etching chemistry.

## V. Conclusion

The full photonic packaging and characterization of the interposer demonstrated a robust, high-density platform for vertical coupling into SiP chips. The femtosecond laser writing process allows for flexibility of mirror arrangements and waveguide routing, shown by the multiple interposer designs that were tested. In the end, an optimized design was selected and fully packaged into a high-density circuit with -5.0 dB average single-pass loss, exploiting the full 3D volume of the glass substrate to weave 40 channels from a 2D MCF array onto a grid of SiP gratings. The continued development of glass circuits promises improved channel numbers and further densification of interconnects, impacting the interconnectivity bottleneck that the integrated optics and telecommunication industries are facing at this moment.